\documentclass[prd,superscriptaddress,twocolumn,eqsecnum,showpacs]{revtex4}
\usepackage[dvips]{graphicx}
\usepackage{color}
\usepackage{latexsym}
\usepackage{amsmath}
\usepackage{amssymb}
\newcommand{\pet}{\pi_k}
\newcommand{\dpet}{\dot \pi_k}
\newcommand{\ddpet}{\ddot \pi_k}
\newcommand{\omepi}{\omega}
\newcommand{\rd}{\mathrm{d}}               % differential
\newcommand{\defby}{\equiv}                % means "defined by"
\newcommand{\bK}{\mathbf{K}}               % bold K
\newcommand{\bP}{\mathbf{P}}               % bold p (polarization)
\newcommand{\be}{\hat{\mathbf{e}}}         % bold hat e (unit vector)
\newcommand{\bsigma}{\boldsymbol{\sigma}}  % bold sigma
\newcommand{\calN}{\mathcal{N}}            % calligraphic N
\newcommand{\oneplusone}{(1+1)}            % ( 1 + 1 )
\newcommand{\twoplusone}{(2+1)}            % ( 2 + 1 )
\newcommand{\threeplusone}{(3+1)}          % ( 3 + 1 )
\newcommand{\Diag}[1]{\mathrm{diag}( \, #1 \, )}   % Diag
\newcommand{\Expect}[1]%                           % Expect
   {\ensuremath{\langle \, #1 \,  \rangle}}
\newcommand{\Comm}[2]%                             % Comm
   {\ensuremath{[ \, #1, #2 \, ]}}
\newcommand{\AntiComm}[2]%                         % AntiComm
   {\ensuremath{\{ \, #1, #2 \, \}}}
\newcommand{\Tr}[1]{\mathrm{Tr} [ \, #1 \, ]}      % Tr
\newcommand{\Exp}[1]{\exp [ \, #1 \, ]}            % Exp
\newcommand{\Intk}[1]{\int_{-\infty}^{+\infty}%
   [\mathrm{d} #1] \, }                % Int k
\newcommand{\Ket}[1]{ | \, #1 \, \rangle }         % ket vector

\begin{document}

%
%%%%%%%%%%%%%%%%%%%%%%%%%%%%%%%%%%%%%%%%%%%%%%%%%%%%%%%%%%%%%%%%%%%%%%
%

% titlepage
%
\preprint{LA-UR-08-XXXX}
\title[\oneplusone-dimensional QED -- backreaction revisited]
   {Fermion pair production in QED and the backreaction problem \\
    in (1+1)-dimensional boost-invariant coordinates revisited}

\author{Bogdan Mihaila}
\email{bmihaila@lanl.gov}
\affiliation{%Materials Science and Technology Division,
   Los Alamos National Laboratory,
   Los Alamos, NM 87545}

\author{John F. Dawson}
\email{john.dawson@unh.edu}
\affiliation{
   Department of Physics,
   University of New Hampshire,
   Durham, NH 03824}

\author{Fred Cooper}
\email{cooper@santafe.edu}
\affiliation{National Science Foundation,
   4201 Wilson Blvd.,
   Arlington, VA 22230}
\affiliation{Santa Fe Institute,
   Santa Fe, NM 87501}

%%%\date{\today, \now \ EST}

\begin{abstract}
We study two different initial conditions for fermions for the problem of pair production of fermions coupled to a classical electromagnetic field with backreaction in \oneplusone\ boost-invariant coordinates.  Both of these conditions are consistent with fermions initially in a vacuum state.  We present results for the proper time evolution of the electric field $E$, the current $J$, the matter energy density $\varepsilon$, and the pressure $p$ as a function of the proper time for these two cases.  We also determine the interpolating number density as a function of the proper time.  We find that when we use a ``first order adiabatic'' vacuum initial condition or a ``free field'' initial condition for the fermion field, we obtain essentially similar behavior for physically measurable quantities.  The second method is computationally simpler, it is twice as fast and involves half the storage required by the first method.
\end{abstract}

\maketitle

%
%%%%%%%%%%%%%%%%%%%%%%%%%%%%%%%%%%%%%%%%%%%%%%%%%%%%%%%%%%%%%%%%%%%%%%
%

% \tableofcontents
% \newpage

%
%%%%%%%%%%%%%%%%%%%%%%%%%%%%%%%%%%%%%%%%%%%%%%%%%%%%%%%%%%%%%%%%%%%%%%
%

\section{Introduction}
\label{s:intro}

Particle production from strong fields has a long history starting with Schwinger's classic paper \cite{r:Schwinger:1951fk}. A detailed history of this subject can be found in two recent reviews \cite{ref:Dunne06,CooperDawsonMihaila0806.1249}.  One of the many applications of pair production has been as a model for particle production in the central rapidity region following a relativistic heavy ion collision. Following such a collision, there is experimental evidence that the production of particles is  ``boost-invariant''~\cite{r:CFSprd75,r:Bjorken:1983zr} which leads to measurable quantities such as energy densities being functions of the fluid proper time alone. The initial conditions we want to study for this problem are that the number of pairs starts out zero and that the initial induced current in the Maxwell (backreaction) equation for the electric field is also zero. We then want to study the proper time evolution of the expectation value of the energy density, pressure, and current of the produced particles and the evolution of the electric field.

In an earlier paper on this topic~\cite{r:Cooper:1993uq} one particular set of initial conditions consistent with having no pairs of particles produced before the collision at initial proper time, $\tau=\tau_0$, led to the need for doubling the number of fermion solutions in order to start with zero induced current in the Maxwell equation for the electric field. In the paper by Cooper \emph{et al.}~\cite{r:Cooper:1993uq}, two sets of solutions for the second-order squared Dirac equation were used in order to satisfy the desired initial conditions of having zero initial current.  Similar results were presented in a 1992 paper by Kluger \emph{et al.}~\cite{r:Kluger:1992fk} for the Cartesian case.  In those papers, the initial conditions were taken to correspond to a first-order adiabatic approximation to the second-order Dirac equation, which forced them to average over two different solutions of the Dirac equation so that the current vanished at $\tau = \tau_0$. In the present work, we consider a slightly different initial state, namely  approximate free fields for the fermions, which automatically leads to a zero current at $\tau=\tau_0$.  This initial condition was used earlier by Cooper and Savage~\cite{r:CS02} in their study of the dynamics of the chiral phase transition in the \twoplusone\ dimensional Gross-Neveu model.  The free  field initial condition does not require doubling the number of solutions as did the adiabatic choice.  We compare the  evolution of the problem for both initial conditions and show that at short to moderate times they are equivalent and are slightly different at very late times.

This semi-classical approximation to the initial value QED problem  describes the fermions as a quantum field but treats the electric field classically.  The current used in Maxwell's equation is calculated using the vacuum expectation value of the quantum Dirac current.  As discussed in previous papers~\cite{r:CHKMPAprd94}, this approximation is equivalent to the first term in a large-N approximation to N-QED where there are N flavors of fermions present.

The method we use for numerically solving this problem is a shooting method to numerically step out solutions of the equations from initial conditions.  An adiabatic analysis of the form of the solutions is used to determine the behavior of the solutions at large momentum and to isolate divergences and perform renormalization as well as to choose appropriate initial states for fermions that are appropriate vacuum states.

We study here the problem in \oneplusone\ boost-invariant coordinates.  This kinematic situation is related to the kinematics of the early phase of plasma evolution following a relativistic heavy ion collision with the electric field a simplification for the semiclassical chromoelectric field expected to be produced in that situation.  We study \oneplusone\ dimensions for simplicity here, where charge renormalization is finite.  However, the same methods of solution used here can be applied to the case of \threeplusone\ dimensions in both Cartesian and boost-invariant coordinates.  We will present results for \threeplusone\ dimensions for QED and QCD elsewhere.

This paper is organized as follows: In Sec.~\ref{s:notations} we review briefly the equations we will need to solve for QED in \oneplusone\ dimensional boost-invariant coordinates (for a detailed derivation, see e.g.\ Refs.~\onlinecite{CooperDawsonMihaila0806.1249} and~\onlinecite{ref:CooperDawsonMihaila06}).  In Secs.~\ref{s:Maxwell} and~\ref{s:adiabaticexp} we discuss the backreaction equation and the adiabatic expansion and charge renormalization, whereas in Sec.~\ref{s:energymomentum} we review the calculation of the energy-momentum tensor. The two types of initial conditions are introduced in Sec.~\ref{s:initialconditions}. We present results of our numerical simulations in Sec.~\ref{s:results} and conclude in Sec.~\ref{s:conclusions}.

%
%%%%%%%%%%%%%%%%%%%%%%%%%%%%%%%%%%%%%%%%%%%%%%%%%%%%%%%%%%%%%%%%%%%%%%
%

\section{Notation and equations}
\label{s:notations}

In our simplified kinematics, in \oneplusone\ dimensions, we choose the longitudinal axis of the collision to be the $z$-axis. Then, in the Cartesian frame, we want to solve the set of equations:
\begin{equation}\label{bi.e:diracI}
   \bigl \{ \,
      \gamma^{a} \, [ \, i \partial_a - e \, A_{a}(\xi) \, ]
      -
      m \,
   \bigr \} \, \hat{\psi}(\xi)
   =
   0 \>,
\end{equation}
where $\xi$ is shorthand for the Cartesian pair $(t,z)$, $\hat{\psi}(\xi)$ is a fermi field satisfying the anti-commutation relation:
\begin{equation}\label{bi.e:fermianticomm}
   \AntiComm{\hat{\psi}_{\alpha}^{\phantom\dagger}(z,t)}
            {\hat{\psi}_{\alpha}^{\dagger}(z',t)}
   =
   \delta_{\alpha,\alpha'} \,
   \delta(z - z') \>,
\end{equation}
and $A_a(\xi)$ is a classical field satisfying Maxwell's equations:
\begin{equation}\label{bi.e:maxwellI}
   \partial_a \, F^{ab}(\xi)
   =
   J^b(\xi) \>,
   \quad
   F^{ab}(\xi)
   =
   \partial^a A^b(\xi) - \partial^b A^a(\xi) \>.
\end{equation}
The current is given by:
\begin{equation}\label{bi.e:JdefI}
   J^b(\xi)
   =
   \frac{e}{2} \,
   \Expect{ \Comm{ \hat{\bar{\psi}}(\xi) }
                 { \gamma^a \, \hat{\psi}(\xi) } } \>.
\end{equation}
The $\gamma$-matrices satisfy $\AntiComm{ \gamma^a }{ \gamma^b }=2\, \eta^{a,b}$ and are given by:
\begin{equation*}
   \gamma^0
   =
   \begin{pmatrix}
      1 & 0 \\
      0 & -1
   \end{pmatrix} \>,
   \>\>
   \gamma^3
   =
   \begin{pmatrix}
      0 & 1 \\
     -1 & 0
   \end{pmatrix} \>,
   \>\>
   \gamma^5
   =
   \gamma^0 \gamma^3
   =
   \begin{pmatrix}
      0 & 1 \\
      1 & 0
   \end{pmatrix} \>.
\end{equation*}

Boost-invariant coordinates $x^\mu = ( \, \tau, \eta \, )$ are defined by:
\begin{equation}\label{bi.e:tauetadef}
   t
   =
   \tau \, \cosh{\eta} \>,
   \qquad
   z
   =
   \tau \, \sinh{\eta} \>.
\end{equation}
The connection between the Cartesian frame ($d\xi^a$), and the boost-invariant frame  ($dx^{\mu}$) is described by a vierbein matrix $V^{a}{}_{\mu}(x)$, given by:
\begin{gather}
   \rd \xi^{a}
   =
   V^{a}{}_{\mu}(x) \, \rd x^{\mu} \>,
   \qquad
   \partial_{\mu}
   =
   V^{a}{}_{\mu}(x) \, \partial_{a} \>,
   \label{bi.e:Vdef} \\
   V^{a}{}_{\mu}(x)
   \defby
   \frac{\partial \xi^{a}}{\partial x^{\mu}}
   =
   \begin{pmatrix}
      \cosh \eta, & \tau \sinh \eta \\
      \sinh \eta, & \tau \cosh \eta
   \end{pmatrix} \>,
   \notag
\end{gather}
and its inverse:
\begin{gather}
   \rd x^{\mu}
   =
   V^{\mu}{}_{a}(x) \, \rd \xi^{a} \>,
   \quad
   \partial_{a}
   =
   V^{\mu}{}_{a}(x) \, \partial_{\mu} \>,
   \label{bi.e:Vinvdef} \\
   V^{\mu}{}_{a}(x)
   \defby
   \frac{\partial x^{\mu}}{\partial \xi^{a}}
   =
   \begin{pmatrix}
      \cosh \eta, & - \sinh \eta \\
      - \sinh \eta / \tau, & \cosh \eta / \tau
   \end{pmatrix} \>.
   \notag
\end{gather}
The $\gamma$-matrices in this frame are denoted by a tilde: $\tilde{\gamma}^{\mu}(x) \defby V^{\mu}{}_{a}(x) \, \gamma^{a}$, and satisfy:
\begin{equation}\label{bi.ae.e:gammamugammanu}
   \AntiComm{ \tilde{\gamma}^{\mu}(x) }{ \tilde{\gamma}^{\nu}(x) }
   =
   2 \, g^{\mu\nu}(x) \>,
   \quad
   g^{\mu\nu}(x)
   =
   \Diag{1, -1/\tau^2} \>.
\end{equation}
Dirac's equation \eqref{bi.e:diracI} becomes in this frame:
\begin{equation}\label{bi.e:diracII}
   \bigl [ \,
      \tilde{\gamma}^{\mu}(x) \,
      ( \,
         i \, \partial_{\mu} - e \, A_{\mu}(x) \,
      )
      -
      m \,
   \bigr ] \,
   \hat{\psi}(x)
   =
   0 \>,
\end{equation}
where the fermi field obeys the anti-commutation relation:
\begin{equation}\label{bi.e:psifieldanticomm}
   \AntiComm{ \hat{\psi}_{\alpha}^{\phantom(}(\tau,\eta) }
            { \hat{\bar{\psi}}_{\alpha'}^{\phantom(}(\tau,\eta') }
   =
   \tilde{\gamma}^{\tau}_{\alpha,\alpha'}(\eta) \,
   \delta( \eta - \eta' ) / \tau \>.
\end{equation}
However, it is much simpler to make a similarity transformation to a system of coordinates where the vierbein becomes diagonal~\cite{ref:CooperDawsonMihaila06}.
In this rotated system, the $\gamma$-matrices are denoted by a bar:
\begin{equation}\label{bi.e:gammabar}
   S^{-1}(\eta) \,
   \tilde{\gamma}^{\mu}(x) \,
   S(\eta)
   =
   \bar{\gamma}^{\mu}(\tau) \>,
\end{equation}
where the matrix $S(\eta)$ is given by:
\begin{equation}\label{bi.e:Sdef}
   S(\eta)
   =
   \Exp{ \eta \, \gamma^5 / 2 }
   =
   \cosh( \eta/2 ) + \gamma^5 \, \sinh( \eta /2 ) \>.
\end{equation}
The $\bar{\gamma}^{\mu}(\tau)$ matrices are given explicitly by:
\begin{equation}\label{bi.e:Vbardefs}
   \bar{\gamma}^{\tau}
   =
   \gamma^{0} \>,
   \qquad
   \bar{\gamma}^{\eta}(\tau)
   =
   \gamma^3 / \tau \>,
\end{equation}
So if we define a new fermi field $\hat{\phi}(x)$ by:
\begin{equation}\label{bi.e:psitophi}
   \hat{\psi}(x)
   =
   S(\eta) \,
   \hat{\phi}(x) / \sqrt{\tau} \>,
\end{equation}
Dirac's equation \eqref{bi.e:diracII} becomes:
\begin{equation}\label{bi.e:diracIII}
   \bigl [ \,
      i \,
      \bar{\gamma}^{\mu}(\tau) \,
      \nabla_{\mu}
      -
      m \,
   \bigr ] \,
   \hat{\phi}(\tau,\eta) / \sqrt{\tau}
   =
   0 \>,
\end{equation}
where $\nabla_{\mu} = \partial_{\mu} + \Pi_{\mu}(x) + i e \, A_{\mu}(x)$, with $\Pi_{\mu}(x) = S^{-1}(\eta) \, ( \, \partial_{\mu} S(\eta) \, )$, is the covariant derivative.  Here $\hat{\phi}(x)$ obeys the simpler anti-commutation relation:
\begin{equation}\label{bi.e:phifieldanticomm}
   \AntiComm{ \hat{\phi}_{\alpha}^{\phantom\dagger}(\tau,\eta) }
            { \hat{\phi}_{\alpha'}^{\dagger}(\tau,\eta') }
   =
   \delta_{\alpha,\alpha'}^{\phantom\dagger} \,
   \delta( \eta - \eta' )  \>.
\end{equation}
For our case, the only non-vanishing $\Pi_{\mu}(x)$ is for $\Pi_{\eta} = \gamma^5 / 2$.  In the boost-invariant frame, we work in the temporal gauge and choose $A_{\mu}(x) = (\, 0, -A(\tau) \,)$.  That is $A(\tau)$ as the negative of the covariant component in the boost-invariant frame.  So \eqref{bi.e:diracIII} simplifies to:
\begin{equation}\label{bi.e:diracIV}
   \bigl \{ \,
   i \, \gamma^0 \,
   \partial_\tau
   +
   \gamma^3 \,
   \bigl [ \,
      i \, \partial_{\eta}
      +
      e \, A(\tau) \,
   \bigr ] / \tau
   -
   m \,
   \bigr \} \, \hat{\phi}(\tau,\eta)
   =
   0 \>.
\end{equation}

We now expand the field $\hat{\phi}(\tau,\eta)$ in a fourier series given by:
\begin{equation}\label{bi.e:phiexpand}
   \hat{\phi}(\tau,\eta)
   =
   \Intk{k} \sum_{\lambda=\pm}
   \hat{A}_{k}^{(\lambda)} \,
   e^{i k \eta} \,
   \phi_{k}^{(\lambda)}(\tau) \>,
\end{equation}
where we have introduced the notation $[\mathrm{d}k] = \mathrm{d}k/(2\pi)]$. Here, $\hat{A}_{k}^{(\lambda)}$ are mode operators and $\phi_{k}^{(\lambda)}(\tau)$ are two independent mode functions satisfying the equation:
\begin{equation}\label{bi.e:diracV}
   \bigl [ \,
   i \, \gamma^0 \,
   \partial_\tau
   -
   \gamma^3 \,
   \pet(\tau)
   -
   m \,
   \bigr ] \, \phi_k^{(\lambda)}(\tau)
   =
   0 \>,
\end{equation}
where
\begin{equation}
   \pet(\tau) = \frac{1}{\tau} \ [ \, k - e \, A(\tau) \, ]
   \>.
\end{equation}
It is now useful to add and subtract the upper and lower components of the spinor $\phi_k^{(\lambda)}$ by writing:
\begin{equation}\label{bi.e:rot}
    \phi_k^{(\lambda)}(\tau)
    =
    U \, F_k^{(\lambda)}(\tau) \>,
    \>\>\text{with}\>\>
    U
    =
    \frac{1}{\sqrt{2}}
    \begin{pmatrix}
       1 & 1 \\
       1 & -1
    \end{pmatrix} \>.
\end{equation}
Here $U^{\dagger} = U^{-1} = U^{T}$.  Then $F_k^{(\lambda)}(\tau)$ satisfies an equation of Hamiltonian form:
\begin{equation}\label{bi.e:Fequ}
   i \, \partial_\tau \, F_k^{(\lambda)}(\tau)
   =
   H(\tau) \, F_k^{(\lambda)}(\tau) \>,
\end{equation}
with
\begin{equation}\label{bi.de.e:Hdef}
   H(\tau)
   =
   \begin{pmatrix}
      \pet(\tau) & m \\
      m & - \pet(\tau)
   \end{pmatrix}
   =
   \bK_k(\tau) \cdot \bsigma \>.
\end{equation}
Here $\bK_k(\tau)$ is a vector defined in an abstract space $\mathcal{R}$ with unit vectors $(\be_1,\be_2,\be_3)$ and given by:
\begin{equation}\label{bi.e:kvecdef}
   \bK_k(\tau)
   =
   m \, \be_1
   +
   \pet(\tau) \, \be_3 \>.
\end{equation}
We can introduce the $2 \times 2$ dimensional density matrix $\rho_k(\tau)$ and a ``polarization'' vector $\bP_k(\tau)$ in $\mathcal{R}$ with the definitions:
\begin{equation}\label{bi.e:rhobPdefs}
   \rho_k^{(\lambda)}(\tau)
   =
   F_k^{(\lambda)}(\tau) \, F_k^{(\lambda)\,\dagger}(\tau)
   =
   \frac{1}{2} \,
   ( \, 1 + \bP_k^{(\lambda)}(\tau) \cdot \bsigma \, ) \>.
\end{equation}
Then from \eqref{bi.e:Fequ}, the polarization vector $\bP_k^{(\lambda)}(\tau)$ obeys the vector equation of motion:
\begin{equation}\label{bi.e:Peom}
   \partial_\tau \, \bP_k^{(\lambda)}(\tau)
   =
   2 \, \bK_k(\tau) \times \bP_k^{(\lambda)}(\tau) \>.
\end{equation}
Since $H(\tau)$ in Eq.~\eqref{bi.de.e:Hdef} is hermitian, $F_k^{(\lambda)}(\tau)$ satisfies a conservation equation:
\begin{equation}\label{bi.e:conseqs}
   \partial_\tau \,
   [ \, F_k^{(\lambda)\,\dagger}(\tau) \, F_k^{(\lambda')}(\tau) \, ]
   =
   0 \>.
\end{equation}
So if we choose the two spinors to be orthonormal at $\tau = \tau_0$, they remain orthonormal for all $\tau$.  In Sec.~\ref{s:initialconditions} we show how to do this.  So we can assume that these spinors are orthonormal and complete for all $\tau$:
\begin{subequations}\label{bi.e:orthocomp}
\begin{align}
   F_{k}^{(\lambda)\,\dagger}(\tau) \,
   F_{k}^{(\lambda')}(\tau)
   &=
   \delta_{\lambda,\lambda'} \>,
   \label{bi.de.e:ortho} \\
   \sum_{\lambda=\pm}
   F_{k}^{(\lambda)}(\tau) \,
   F_{k}^{(\lambda)\,\dagger}(\tau)
   &=
   1 \>.
   \label{bi.de.e:complete}
\end{align}
\end{subequations}
Probability conservation also requires that the polarization vector $\bP_{k}^{(\lambda)}(\tau)$ for both of these solutions to remain on the unit sphere for all time $\tau$.  So to summarize, the fermi field can be written as:
\begin{equation}\label{bi.e:psitoF}
   \hat{\psi}(\tau,\eta)
   =
   S(\eta) \, U \,
   \hat{F}(\tau,\eta) / \sqrt{\tau} \>,
\end{equation}
where the field $\hat{F}(\tau,\eta)$ obeys the anti-commutation relation:
\begin{equation}\label{bi.e:Ffieldanticomm}
   \AntiComm{ \hat{F}_{\alpha}^{\phantom\dagger}(\tau,\eta) }
            { \hat{F}_{\alpha'}^{\dagger}(\tau,\eta') }
   =
   \delta_{\alpha,\alpha'}^{\phantom\dagger} \,
   \delta( \eta - \eta' )  \>.
\end{equation}
and is expanded in terms of the spinors $F_{k}^{(\lambda)}(\tau)$ which satisfy \eqref{bi.e:Fequ}:
\begin{equation}\label{bi.e:Fexpansion}
   \hat{F}(\tau,\eta)
   =
   \Intk{k} \sum_{\lambda=\pm}
   \hat{A}_{k}^{(\lambda)} \,
   e^{i k \eta} \,
   F_{k}^{(\lambda)}(\tau) \>.
\end{equation}

We can use this orthogonality to invert \eqref{bi.e:Fexpansion} to get:
\begin{equation}\label{bi.e:Aexp}
   \hat{A}_{k}^{(\lambda)}
   =
   \int_{-\infty}^{+\infty} \rd \eta \,
    e^{-i k \eta} \,
    F_{k}^{(\lambda)\,\dagger}(\tau) \,
    \hat{F}(\tau,\eta) \>,
\end{equation}
for any time $\tau$.  Using \eqref{bi.e:Ffieldanticomm}, we then find that the mode operators $\hat{A}_{k,s}^{(\lambda)}$ obey the anti-commutation relation:
\begin{equation}\label{bi.e:Aanticomm}
   \AntiComm{ \hat{A}_{k}^{(\lambda)} }
            { \hat{A}_{k'}^{(\lambda')\,\dagger} }
   =
   ( 2\pi ) \, \delta_{\lambda,\lambda'}^{\phantom(} \, \delta( k - k' ) \>.
\end{equation}
It is traditional to define separate positive and negative energy operators by setting:
\begin{equation}\label{bi.e:tradition}
   \hat{A}_{k}^{(+)}
   =
   \hat{a}_{k}^{\phantom(} \>,
   \qquad\text{and}\qquad
   \hat{A}_{k}^{(-)}
   =
   \hat{b}_{-k}^{\dagger} \>.
\end{equation}
We choose our initial state to be the vacuum with no particle or anti-particle present.  Then:
\begin{equation}\label{bi.e:abvac}
   \hat{a}_{k}^{\phantom(} \, \Ket{0} = 0 \>,
   \qquad\text{and}\qquad
   \hat{b}_{k}^{\phantom(} \, \Ket{0} = 0 \>.
\end{equation}
This means that:
\begin{equation}\label{bi.e:expectcomm}
   \Expect{ \Comm{ \hat{A}_{k}^{(\lambda)\,\dagger} }
                 { \hat{A}_{k'}^{(\lambda')} } }
   =
   - \,
   ( 2\pi ) \, \lambda \, \delta_{\lambda,\lambda'}^{\phantom(} \, \delta( k - k' ) \>,
\end{equation}
a result we will use in the next section.

%
%%%%%%%%%%%%%%%%%%%%%%%%%%%%%%%%%%%%%%%%%%%%%%%%%%%%%%%%%%%%%%%%%%%%%%
%

\section{Maxwell's equation}
\label{s:Maxwell}

Maxwell's equation is given in Cartesian coordinates in Eq.~\eqref{bi.e:maxwellI} with the current given in Eq.~\eqref{bi.e:JdefI}.  For our boost-invariant coordinates, Maxwell's equation reads:
\begin{equation}\label{bi.me.e:maxwellcov}
   \frac{1}{\sqrt{-g}} \,
   \partial_{\mu}
   \bigl [ \,
      \sqrt{-g} \, F^{\mu\nu}(x) \,
   \bigr ]
   =
   J^{\nu}(x) \>,
\end{equation}
where $\sqrt{-g} = \tau$.  Now $A_{\mu} = ( \, 0, - A(\tau) \, )$, so the only non-vanishing elements of the field tensor are:
\begin{equation}\label{bi.me.e:Edef}
   F_{\tau,\eta}(x)
   =
   - F_{\eta,\tau}(x)
   =
   -
   \partial_\tau A(\tau)
   \defby
   \tau \, E(\tau) \,
\end{equation}
This last equation defines what we call the electric field $E(\tau) \defby ( \, \partial_\tau A(\tau) \, ) / \tau$.  Then using the metric $g^{\mu\nu}(x) = \Diag{ 1, -1/\tau^2 }$, we get:
\begin{equation}\label{bi.me.e:Fupper}
   F^{\tau,\eta}(\tau)
   =
   - F^{\eta,\tau}(\tau)
   =
   -
   E(\tau) / \tau \>,
\end{equation}
and Maxwell's equation becomes:
\begin{equation}\label{bi.me.e:maxwellcovII}
   \partial_\tau E(\tau)
   =
   - J(\tau) \>.
\end{equation}
Here we have defined a ``reduced'' current $J(\tau)$ by:
\begin{equation}\label{bi.me.e:jsdef}
\begin{split}
   J(\tau)
   &=
   \frac{e \, \tau}{2} \,
   \Expect{
      \Comm{ \hat{\bar{\psi}}(\eta,\tau) }
           { \tilde{\gamma}^{\eta}(\tau) \,
             \hat{\psi}(\eta,\tau) } }
   \\
   &=
   \frac{e}{2 \, \tau} \,
   \Expect{
      \Comm{ \hat{\phi}^{\dagger}(\eta,\tau) }
      { \gamma^5 \, \hat{\phi}(\eta,\tau) } } \>,
\end{split}
\end{equation}
Using the field expansion \eqref{bi.e:phiexpand} and the expectation value \eqref{bi.e:expectcomm} of the mode operators, we find for the reduced current:
\begin{align}
   J(\tau)
   &=
   \frac{e}{2 \, \tau}
   \Intk{k} \sum_{\lambda=\pm 1}
   \Intk{k'} \sum_{\lambda'=\pm 1}
   \notag \\
   &\times
   e^{i (k - k') \eta} \,
   \bigl [ \,
      \phi_{k}^{(\lambda)\,\dagger}(\tau) \,
      \gamma^5 \,
      \phi_{k'}^{(\lambda')\phantom\dagger}(\tau) \,
   \bigr ] \,
   \Expect{ \Comm{ \hat{A}_{k}^{(\lambda)\,\dagger} }
                 { \hat{A}_{k'}^{(\lambda')} } }
   \notag \\
   &=
   -
   \frac{e}{2 \, \tau}
   \Intk{k} \sum_{\lambda=\pm 1}
   \lambda \,
   \bigl [ \,
      F_{k}^{(\lambda)\,\dagger}(\tau) \,
      \sigma_3 \,
      F_{k}^{(\lambda)\phantom\dagger}(\tau) \,
   \bigr ]
   \notag \\
   &=
   - e
   \Intk{\pet} \> P_3^{(+)}(\pet,\tau) \>.
   \label{bi.me.e:js}
\end{align}
Here we have used the completeness statement \eqref{bi.de.e:complete}
to write the current in terms of positive energy solutions only.  In the last line, we changed integration variables from $k$ to $\pet(\tau)$, using $\rd \pet = \rd k / \tau$, and defined $\bP(\pet,\tau) \defby \bP_k(\tau)$.  Maxwell's equation \eqref{bi.me.e:maxwellcovII} becomes:
\begin{equation}\label{bi.me.e:maxwellfinal}
   \partial_\tau E(\tau)
   =
   e
   \Intk{\pet} \> P_3^{(+)}(\pet,\tau) \>.
\end{equation}
Recall that $P_3$ is the third component of the polarization vector in the space $\mathcal{R}$.

%
%%%%%%%%%%%%%%%%%%%%%%%%%%%%%%%%%%%%%%%%%%%%%%%%%%%%%%%%%%%%%%%%%%%%%%
%

\section{Adiabatic expansion}
\label{s:adiabaticexp}

The large momentum behavior of the solutions of the Dirac equation can be obtained by looking at the adiabatic expansion of these solutions.  Perhaps the simplest way to do this is from the polarization equation \eqref{bi.e:Peom}.  In order to count powers of time derivatives, we put:
$\partial_\tau \mapsto \epsilon \, \partial_\tau$, and set:
\begin{equation}\label{bi.ae.e:PexpandI}
   \bP_k^{\phantom)}(\tau)
   =
   \bP_k^{(0)}(\tau)
   +
   \epsilon \, \bP_k^{(1)}(\tau)
   +
   \epsilon^2 \, \bP_k^{(2)}(\tau)
   +
   \dotsb
\end{equation}
Substitution of this into Eq.~\eqref{bi.e:Peom} and equating powers of $\epsilon$ give the results:
\begin{subequations}\label{bi.ae.e:PexpandII}
\begin{align}
   \bP_k^{(0)}
   &=
   \frac{\bK_k}{\omepi} \>,
   \label{bi.ae.e:PexpA} \\
   \bP_k^{(1)}
   &=
   \frac{ \dot{\bK}_k \times \bK_k }{ 2 \, \omepi^3 } \>,
   \label{bi.ae.e:PexpB} \\
   \bP_k^{(2)}
   &=
   \frac{ 3 \, ( \dot{\bK}_k \cdot \bK_k ) \,  \dot{\bK}_k
          - \omepi^2 \, \ddot{\bK}_k }
        { 4 \, \omepi^5 }
   +
   \calN_k \, \bK \,
   \>,
   \label{bi.ae.e:PexpC}
\end{align}
\end{subequations}
where $\omepi = \sqrt{ \pet^2 + m^2 }$ and
\begin{equation}\label{bi.ae.e:canNdef}
   \calN_k
   =
   -
   \frac{1}{8} \, \frac{\dot{\pet}^2}{\omepi^5}
   +
   \frac{1}{4} \, \frac{\pet \, \ddot{\pet}}{\omepi^5}
   -
   \frac{5}{8} \, \frac{\pet^2 \, \dot{\pet}^2}{\omepi^7} \>.
\end{equation}
We have suppressed the $\tau$ dependence here of these quantities.  The dot denotes a partial derivative with respect to $\tau$.  Explicitly, we find:
\begin{align}
   P_{1}
   &=
   \frac{m}{\omepi}
   +
   \epsilon^2 \, m \,
   \Bigl ( \,
      -
      \frac{1}{8} \,
      \frac{\dpet^2}{\omepi^5}+
      \frac{1}{4} \,
      \frac{ \pet \, \ddpet}{\omepi^5}
      -
      \frac{5}{8} \,
      \frac{\pet^2 \, \dpet^2}{\omepi^7} \,
   \Bigr )
   +
   \dotsb
   \notag \\
   P_{2}
   &=
   \epsilon \, m \,
   \frac{\dpet}{2 \, \omepi^3}
   +
   \dotsb
   \notag \\
   P_3
   &=
   \frac{\pet}{\omepi}
   -
   \epsilon^2 \, m^2 \,
   \Bigl ( \,
      \frac{1}{4} \,
      \frac{ \ddpet }{ \omepi^5 }
      -
      \frac{5}{8} \,
      \frac{ \pet \, \dpet^2 }{ \omepi^7 } \,
   \Bigr )
   +
   \dotsb
   \label{bi.ae.e:PxPyPz}
\end{align}
So setting $\epsilon \rightarrow 1$, Maxwell's equation \eqref{bi.me.e:maxwellfinal} becomes:
\begin{equation}\label{bi.ae.e:Maxwellad}
   \dot{E}(\tau)
   =
   e
   \Intk{\pet}
   \biggl [
      \frac{\pet}{\omepi}
      -
      m^2 \,
      \Bigl (
         \frac{1}{4} \,
         \frac{ \ddpet }{ \omepi^5 }
         -
         \frac{5}{8} \,
         \frac{ \pet \, \dpet^2 }{ \omepi^7 }
      \Bigr )
   \biggr ]
   +
   \dotsb
\end{equation}
All terms odd in $\pet$ vanish by symmetric integration.  After integration, Eq.~\eqref{bi.ae.e:Maxwellad} becomes:
\begin{equation}\label{bi.ae.e:MaxwelladII}
   \dot{E}(\tau)
   =
   -
   \frac{ e^2 }{ 6 \pi \, m^2} \, \dot{E}(\tau)
   + J^\mathrm{sub}(\tau) \>.
\end{equation}
Here, the first term corresponds to finite charge renormalization in \oneplusone\ dimensions and can be brought over to the left hand side of the equation.  The current $J^\mathrm{sub}(\tau)$ is explicitly finite by power counting and is initially zero.

An adiabatic expansion of the Dirac equation can also be carried out from solutions of the second-order form of the Dirac equation.  In Section~\ref{ss:twofield} below, we show that this gives the same result as in Eqs.~\eqref{bi.ae.e:PxPyPz}.

%
%%%%%%%%%%%%%%%%%%%%%%%%%%%%%%%%%%%%%%%%%%%%%%%%%%%%%%%%%%%%%%%%%%%%%%
%

\section{Energy-momentum tensor}
\label{s:energymomentum}

In the boost-invariant coordinate system, the average value of the total energy-momentum tensor is given by Eqs.~(4.1) and (4.2) of Ref.~\onlinecite{r:Cooper:1993uq}, and is the sum of two terms:
\begin{equation}\label{bi.em.e:Tmunu}
   T_{\mu\nu}
   =
   T_{\mu\nu}^{\text{matter}}
   +
   T_{\mu\nu}^{\text{field}}
   =
   \Diag{ \mathcal{E}, \tau^2 \, \mathcal{P} } \>,
\end{equation}
where
\begin{subequations}\label{bi.em.e:Tmf}
\begin{align}
   T_{\mu\nu}^{\text{matter}}
   &=
   \frac{1}{4}
   \bigl \langle
      \Comm{ \hat{\bar{\psi}}(x) }
           { \tilde{\gamma}_{(\mu}^{\phantom\ast}(x) \,
             ( i \, D_{\nu)}^{\phantom\ast} \, \hat{\psi}(x) ) }
      +
      \text{h.c.} \,
   \bigr \rangle
   \label{em.e:Tmatter} \\
   T_{\mu\nu}^{\text{field}}
   &=
   g_{\mu\nu} \, \frac{1}{4} \, F^{\alpha\beta} F_{\alpha\beta}
   +
   F_{\mu\alpha} \, g^{\alpha\beta} \,F_{\beta\nu} \>.
   \label{em.e:Tfield}
\end{align}
\end{subequations}
Here $D_{\mu} = \partial_{\mu} + i e \, A_{\mu}(x)$ and the subscript notation $(\mu,\nu)$ means to symmetrize the term.  From our definitions in Section~\ref{s:Maxwell} and Eq.~\eqref{bi.me.e:Edef}, the field part of the energy-momentum tensor is given by:
\begin{equation}\label{bi.em.e:fieldEMtensor}
   T_{\mu\nu}^{\text{field}}
   =
   \Diag{ E^2/2, - \tau^2 \, E^2/2 } \>.
\end{equation}
We denote the matter part of the energy-momentum tensor as:
\begin{equation}\label{matter:Tmunu}
   T_{\mu\nu}^{\text{matter}}
   =
   \Diag{ \varepsilon, \tau^2 \, p } \>.
\end{equation}
For the matter field, we first note that $D_{\nu} \, \hat{\psi}(x) = S(x) \, \nabla_{\nu} \, \hat{\phi}(x) / \sqrt{\tau}$, where $\nabla_{\nu}$ is the covariant derivative defined below Eq.~\eqref{bi.e:diracIII}.  For the $T_{\tau\tau}=\varepsilon(\tau)$ component, $\nabla_{0} = \partial_\tau$, and using the field expansion \eqref{bi.e:phiexpand}, Eqs.~\eqref{bi.e:rot},\eqref{bi.e:Fequ}, and \eqref{bi.e:expectcomm}, we find:
\begin{align}
   \varepsilon(\tau)
   &=
   -
   \frac{1}{2 \tau}
   \Intk{k} \sum_{\lambda} \lambda \,
   \Tr{ \rho_k^{(\lambda)}(\tau) \, H(\tau) }
   \notag \\
   &=
   -
   \frac{1}{\tau} \Intk{k} \,
   \Tr{ \rho_k^{(+)}(\tau) \, H(\tau) }
   \notag \\
   &=
   -
   \Intk{\pet} \>
   \bK(\pet) \cdot \bP^{(+)}(\pet,\tau) \>.
   \label{bi.em.e:Ttt}
\end{align}
For the $T_{\eta\eta}=\tau^2 p(\tau)$ component, $\nabla_{\eta} = \partial_{\eta} - i e A(\tau) + \gamma^5/2$.  Following similar steps to the preceding calculation, we find:
\begin{align}
   p(\tau)
   &=
   -
   \frac{1}{2\tau}
   \Intk{k} \sum_{\lambda} \lambda \, \pet(\tau) \,
   \Tr{ \rho_k^{(\lambda)}(\tau) \, \sigma_3 }
   \notag \\
   &=
   -
   \frac{1}{\tau}
   \Intk{k} \, \pet(\tau) \,
   \Tr{ \rho_k^{(+)}(\tau) \, \sigma_3 }
   \notag \\
   &=
   -
   \Intk{\pet} \> \pet \, P_3^{(+)}(\pet,\tau) \>.
   \label{bi.em.e:Tee}
\end{align}
So from \eqref{bi.em.e:Tmunu},
\begin{subequations}\label{bi.em.e:EPvalues}
\begin{align}
   \mathcal{E}
   &=
   -
   \Intk{\pet} \>
   \bK(\pet) \cdot \bP^{(+)}(\pet,\tau)
   +
   \frac{E^2}{2} \>,
   \label{bi.em.e:Evalue} \\
   \mathcal{P}
   &=
   -
   \Intk{\pet} \> \pet \, P_3^{(+)}(\pet,\tau)
   -
   \frac{E^2}{2} \>.
   \label{bi.em.e:Pvalue}
\end{align}
\end{subequations}

The covariant derivative of the energy-momentum tensor in boost-invariant  coordinates is conserved:
\begin{equation}\label{bi.em.e:consI}
   T^{\mu\nu}{}_{;\mu}
   =
   \partial_{\mu} T^{\mu\nu}
   +
   \Gamma^{\mu}_{\mu\sigma} T^{\sigma\nu}
   +
   \Gamma^{\nu}_{\mu\sigma} T^{\mu\sigma}
   =
   0 \>.
\end{equation}
The Christoffel symbols are defined by: $\Gamma^{\lambda}_{\mu\nu}(x) = V^{\lambda}{}_{a}(x) \, ( \partial_{\mu} V^{a}{}_{\nu}(x) )$.
In our case, the non-vanishing symbols are given by:
\begin{equation}\label{bi.em.e:Christoffel}
   \Gamma^{\tau}_{\eta\eta}
   =
   \tau \>,
   \qquad
   \Gamma^{\eta}_{\tau\eta}
   =
   \Gamma^{\eta}_{\eta\tau}
   =
   1 / \tau \>.
\end{equation}
So we find that
\begin{equation}\label{bi.em.e:ConsI}
   \partial_\tau \, T^{\tau\tau}
   +
   T^{\tau\tau} / \tau
   +
   \tau \, T^{\eta\eta}
   =
   0 \>,
\end{equation}
or
\begin{equation}\label{bi.em.e:ConsII}
   \partial_\tau \, ( \tau \mathcal{E} )
   +
   \mathcal{P}
   =
   0 \>.
\end{equation}
Using the equation of motion \eqref{bi.e:Peom} and Maxwell's equation \eqref{bi.me.e:maxwellfinal}, one can show that Eq.~\eqref{bi.em.e:ConsII} is automatically satisfied.

Using Eqs.~\eqref{bi.ae.e:PxPyPz}, the adiabatic expansion for the energy is given by:
\begin{equation}\label{bi.em.e:Eadiab}
   \mathcal{E}
   =
   \biggl (
      1
      +
      \frac{e^2}{6\pi \, m^2}
   \biggr ) \,
   \frac{E^2}{2}
   +
   \frac{1}{24 \pi \, \tau^2}
   -
   \Intk{\pet} 2 \, \omepi
   +
   \dotsb
\end{equation}
We recognize the first term as a finite renormalization of the charge, the second term as a renormalization of the cosmological constant, and the third term as a sum of the zero point energies of pairs  of particles and anti-particles with energy $\omepi(\pet)$.  We subtract these terms from the calculation of the energy and arrive at a finite energy $\mathcal{E}^{\text{sub}}$ given by:
\begin{equation}\label{bi.em.e:Esub}
   \mathcal{E}^{\text{sub}}	
   =
   \frac{E^2}{2}
   +
   \Intk{\pet} \,
   \Bigl [ \,   -
      \bK(\pet) \cdot \bP(\pet,\tau)
      +
      \omepi
      -
      \frac{\dpet^2}{\omepi^5} \,
   \Bigr ]
   \>.
\end{equation}
For the pressure, the adiabatic expansion gives:
\begin{equation}\label{bi.em.e:Padiab}
   \mathcal{P}
   =
   -
   \Bigl (
      1
      +
      \frac{e^2}{6\pi \, m^2}
   \Bigr ) \,
   \frac{E^2}{2}
   -
   \frac{1}{8 \pi \, \tau^2}
   -
   \Intk{\pet} \frac{2\, \pet^2 }{\omepi}
   +
   \dotsb
\end{equation}
Again, the first term renormalizes the charge, the second term in canceled by the cosmological constant term and the third is the usual pressure.  We subtract these terms from the pressure to get:
\begin{align}
   \mathcal{P}^{\text{sub}}	
   =
   -
   \frac{E^2}{2}
   + &
   \Intk{\pet} \,
   \Bigl [ \,
      -
      \pet \, P_3(\pet,\tau)
\label{bi.em.e:Psub}
   \\ \notag & \quad
      +
      \frac{\pet^2}{\omepi}
      -
      m^2 \,
      \Bigl ( \,
         \frac{1}{4}
         \frac{ \pet \, \dpet^2}{\omepi^5}
         -
         \frac{5}{8}
         \frac{\pet^2 \, \dpet^2}{\omepi^7} \,
      \Bigr ) \,
   \Bigr ]
   \>.
\end{align}
Eqs.~\eqref{bi.em.e:Esub} and \eqref{bi.em.e:Psub} are now finite.

%
%%%%%%%%%%%%%%%%%%%%%%%%%%%%%%%%%%%%%%%%%%%%%%%%%%%%%%%%%%%%%%%%%%%%%%
%

\begin{figure}[t!]
   \centering
   \includegraphics[width=0.9\columnwidth]{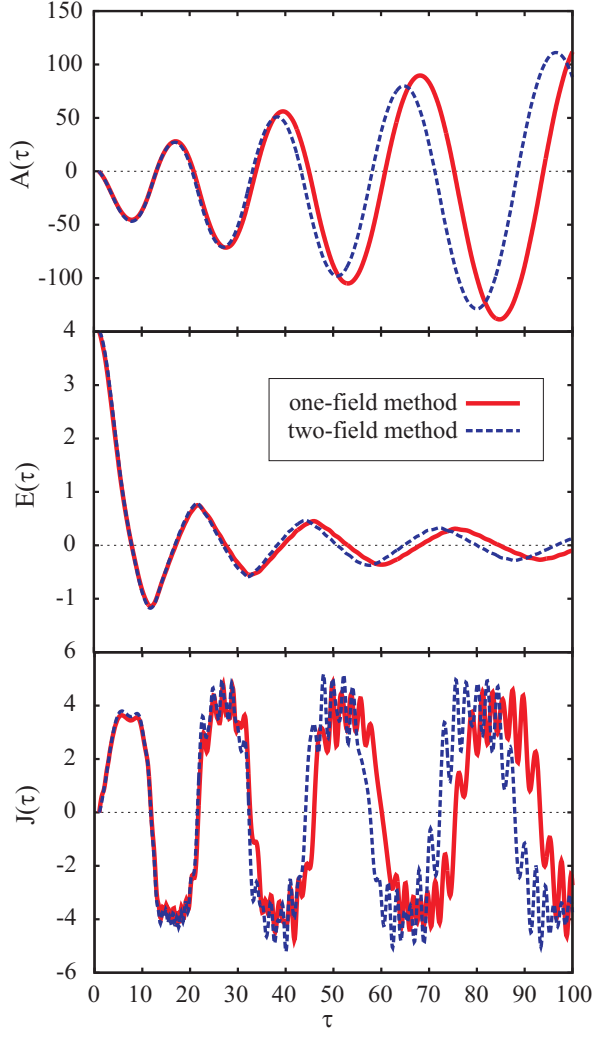}
   \caption{\label{fig1}(Color online)
   Proper-time evolution of the electromagnetic fields and current for
   the one-field and two-field methods described in text.
   Here we choose $m=1$, $A(\tau_0)=0$ and $E(\tau_0)=4$.}
\end{figure}

%
%%%%%%%%%%%%%%%%%%%%%%%%%%%%%%%%%%%%%%%%%%%%%%%%%%%%%%%%%%%%%%%%%%%%%%
%

\section{Initial conditions}
\label{s:initialconditions}

The simplest choice of initial conditions is to find approximate free-field solutions of Eq.~\eqref{bi.e:Fequ} near $\tau = \tau_0$.  This strategy was used in Ref.~\onlinecite{r:CS02}, and automatically provides a zero current at $\tau=\tau_0$.  We call this the ``one-field'' method, and is discussed in Section~\ref{ss:onefield} below.  In previous studies of the backreaction problem by Cooper \emph{et al.}~\cite{r:Cooper:1993uq} adiabatic initial conditions were used which required averaging over two different solutions to the Dirac equation to obtain an zero current at initial proper time $\tau_0$.  We call this the ``two-field'' method.  We discuss this method in Section~\ref{ss:twofield}.

%
%%%%%%%%%%%%%%%%%%%%%%%%%%%%%%%%%%%%%%%%%%%%%%%%%%%%%%%%%%%%%%%%%%%%%%
%

\subsection{One-field method}
\label{ss:onefield}

At $\tau = \tau_0 \equiv 1/m$, $A(\tau_0) = 0$ and $H(\tau_0)$ is given by:
\begin{equation}\label{bi.ic.e:Hzero}
   H(\tau_0)
   =
   m
   \begin{pmatrix}
      k & 1 \\
      1 & - k
   \end{pmatrix} \>.
\end{equation}
So at $\tau \approx \tau_0$, $F_k(\tau)$ obeys the approximate equation of motion:
\begin{equation}\label{bi.ic.e:F0eom}
   i \, \partial_\tau \, F_{0;k}(\tau)
   =
   H(\tau_0) \, F_{0;k}(\tau) \>.
\end{equation}
Writing
\begin{equation}\label{bi.ic.e:F0e}
   F_{0;k}(\tau)
   =
   \tilde{F}_{0;k} \, e^{-i \omepi ( \tau - \tau_0 )} \>,
\end{equation}
We find that $\omepi(\tau_0) = \pm \omega_0$, where $\omega_0 = m \sqrt{k^2 + 1}$.  Positive frequency solutions given by:
\begin{equation}\label{bi.ic.e:F0p}
   \tilde{F}_{0;k}^{(+)}
   =
   \sqrt{ \frac{ \omega_0 + m k }{ 2 \omega_0 } }
   \begin{pmatrix}
      1 \\[2pt]
      \zeta
   \end{pmatrix}
   =
   \begin{pmatrix}
      \cos ( \theta_k / 2 )
      \\[4pt]
      \sin ( \theta_k / 2 )
   \end{pmatrix} \>,
\end{equation}
and negative frequency solutions by:
\begin{equation}\label{bi.ic.e:F0n}
   \tilde{F}_{0;k}^{(-)}
   =
   \sqrt{ \frac{ \omega_0 + m k }{ 2 \omega_0 } }
   \begin{pmatrix}
      - \zeta
      \\[2pt]
      1
   \end{pmatrix}
   =
   \begin{pmatrix}
      - \sin ( \theta_k / 2 )
      \\[4pt]
      \cos ( \theta_k / 2 )
   \end{pmatrix} \>,
\end{equation}
with $\zeta = m / (\omega_0 + m k)$. Here $\sin \theta_k = 1 / \sqrt{k^2 + 1}$ and $\cos \theta_k = k / \sqrt{k^2 + 1}$, with $0 \le \theta_k \le \pi$.  Density matrices for these solutions are given by:
\begin{subequations}\label{bi.ic.e:rho0pm}
\begin{align}
   \rho_k^{(+)}
   &=
   F_{0;k}^{(+)}(\tau) \, F_{0;k}^{(+)\,\dagger}(\tau)
   \label{bi.ic.e:rho0p} \\
   &=
   \begin{pmatrix}
      \cos^2 ( \theta_k / 2 )
      &
      \sin ( \theta_k / 2 ) \, \cos ( \theta_k / 2 )
      \\
      \sin ( \theta_k / 2 ) \, \cos ( \theta_k / 2 )
      &
      \sin^2 ( \theta_k / 2 )
   \end{pmatrix} \>,
   \notag \\
   \rho_k^{(-)}
   &=
   F_{0;k}^{(-)}(\tau) \, F_{0;k}^{(-)\,\dagger}(\tau)
   \label{bi.ic.e:rho0m} \\
   &=
   \begin{pmatrix}
      \sin^2 ( \theta_k / 2 )
      &
      - \sin ( \theta_k / 2 ) \, \cos ( \theta_k / 2 )
      \\
      - \sin ( \theta_k / 2 ) \, \cos ( \theta_k / 2 )
      &
      \cos^2 ( \theta_k / 2 )
   \end{pmatrix} \>,
   \notag
\end{align}
\end{subequations}
and are independent of $\tau$.  The corresponding polarization vectors are also independent of $\tau$ and are given by:
\begin{equation}\label{bi.ic.e:Pol0pm}
   \bP_{0;k}^{(+)}
   =
   \sin \theta_k \, \be_1
   +
   \cos \theta_k \, \be_3
   =
   \frac{\bK_k(\tau_0)}{\omega_0}
   =
   -
   \bP_{0;k}^{(-)} \>,
\end{equation}
The initial spinors are orthogonal and complete:
\begin{subequations}\label{bi.ic.e:orthogcomp}
\begin{align}
   F_{0;k}^{(\lambda)\,\dagger}(\tau) \,
   F_{0;k}^{(\lambda')}(\tau)
   &=
   \delta_{\lambda,\lambda'} \>,
   \\
   \sum_{\lambda=\pm}
   F_{0;k}^{(\lambda)}(\tau) \,
   F_{0;k}^{(\lambda)\,\dagger}(\tau)
   &=
   1 \>.
\end{align}
\end{subequations}
So if we set $F_{k}^{(\lambda)}(\tau_0) = \tilde{F}_{0;k}^{(\lambda)}$ at $\tau = \tau_0$, then the exact solutions remain orthogonal and complete for all $\tau$ and \eqref{bi.e:orthocomp} is satisfied.  As we have seen in Section~\ref{s:Maxwell}, only the positive energy solutions are needed for the backreaction calculation.

%
%%%%%%%%%%%%%%%%%%%%%%%%%%%%%%%%%%%%%%%%%%%%%%%%%%%%%%%%%%%%%%%%%%%%%%
%

\begin{figure}[t!]
   \centering
   \includegraphics[width=0.9\columnwidth]{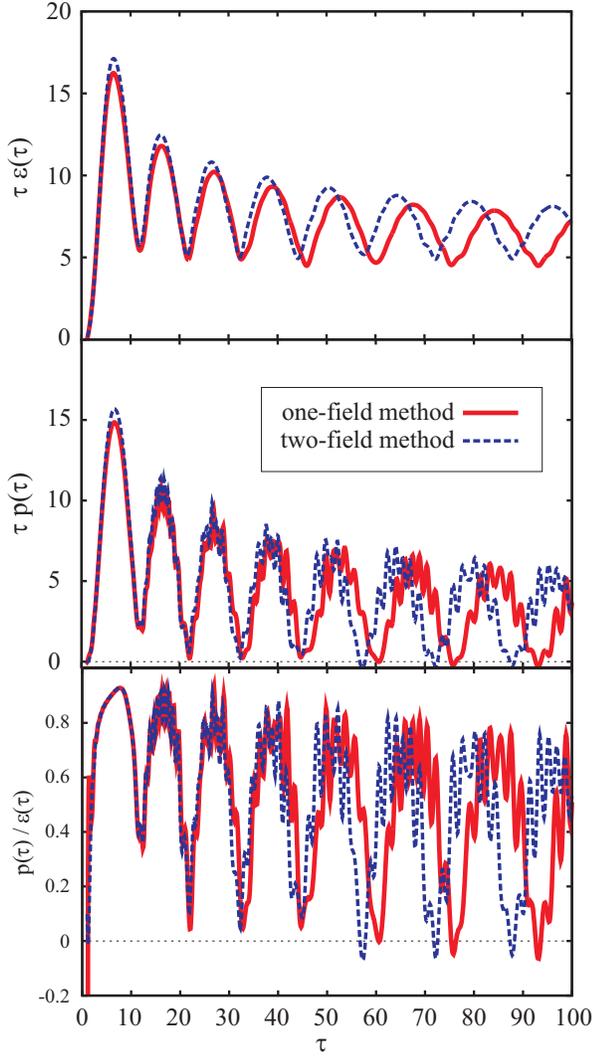}
   \caption{\label{fig2}(Color online)
   Proper-time evolution of the matter components of the renormalized
   energy-momentum tensor for the one-field and two-field methods
   described in text.}
\end{figure}

%
%%%%%%%%%%%%%%%%%%%%%%%%%%%%%%%%%%%%%%%%%%%%%%%%%%%%%%%%%%%%%%%%%%%%%%
%

The initial spinors can serve to define a particle number operator.
Since these initial mode functions form a complete set, we can expand the quantum field in terms of them:
\begin{equation}\label{bi.ic.e:phi0field}
   \hat{F}_{\alpha}(\tau,\eta)
   =
   \Intk{k} \sum_{\lambda} \>
   \hat{A}_{0;k}^{(\lambda)}(\tau) \>
   e^{i k \eta } \>
   F_{0,\alpha;k}^{(\lambda)}(\tau) \>,
\end{equation}
where $\hat{A}_{0;k}^{(\lambda)}(\tau)$ are mode operators for the $F_{0;\alpha;k}^{(\lambda)}(\tau)$ functions, which now depend on time.  Inverting \eqref{bi.ic.e:phi0field}, we find:
\begin{equation}\label{bi.ic.e:A0invert}
   \hat{A}_{0;k}^{(\lambda)}(\tau)
   =
   \int_{-\infty}^{+\infty} \rd x \,
   \sum_{\alpha}
   e^{- i k \eta } \>
   F_{0,\alpha;k}^{(\lambda)\,\ast}(\tau) \,
   \hat{F}_{\alpha}^{\phantom{(}}(\tau,\eta) \>,
\end{equation}
from which we obtain the equal time anti-commutation relation:
\begin{equation}\label{ic.e:A0anticomms}
   \AntiComm{ \hat{A}_{0;k}^{(\lambda)}(\tau) }
            { \hat{A}_{0;k'}^{(\lambda')\,\dagger}(\tau) }
   =
   ( 2\pi ) \, \delta_{\lambda,\lambda'} \, \delta( k - k' ) \>.
\end{equation}
Inserting the expansion \eqref{bi.e:Fexpansion} into the right-hand-side of Eq.~\eqref{bi.ic.e:A0invert}, we can relate the $\hat{A}_{0;k}^{(\lambda)}(\tau)$ mode operators to the $\hat{A}_{k}^{(\lambda)}$ mode operators.  We find:
\begin{equation}\label{ic.e:A0toA}
   \hat{A}_{0;k}^{(\lambda)}(\tau)
   =
   \sum_{\lambda'}
   C_{k}^{(\lambda,\lambda')}(\tau) \,
   \hat{A}_{k}^{(\lambda')} \>,
\end{equation}
where
\begin{equation}\label{ic.e:Cdef}
   C_{k}^{(\lambda,\lambda')}(\tau)
   =
   F_{0;k}^{(\lambda)\,\dagger}(\tau) \,
   F_{k}^{(\lambda')}(\tau) \>.
\end{equation}

Particles are defined in reference to these initial states where a clear distinction between particles and anti-particles can be made.  We define an average phase space number density $n_{k}(\tau)$ by:
\begin{equation}\label{ic.e:ndef}
   n_{k}(\tau)
   =
   \frac{ \rd^2 N(\tau) }{ \rd k \, \rd \eta } \>,
\end{equation}
and is computed using the relation:
\begin{equation}\label{ic.e:ndefI}
   n_{k}(\tau) \,
   ( 2\pi ) \, \delta( k - k' )
   =
   \Expect{ \hat{A}_{0;k}^{(+)\,\dagger}(\tau) \,
            \hat{A}_{0;k'}^{(+)}(\tau) } \>.
\end{equation}
Inserting \eqref{ic.e:A0toA} into \eqref{ic.e:ndefI}, and using
\begin{equation}\label{ic.e:expectAA}
   \Expect{ \hat{A}_{k}^{(\lambda)\,\dagger} \,
            \hat{A}_{k'}^{(\lambda')} }
   =
   \delta_{\lambda,-} \, \delta_{\lambda',-} \,
   ( 2\pi ) \, \delta( k - k' ) \>,
\end{equation}
we find:
\begin{equation}\label{ic.e:nresult}
\begin{split}
   n_{k}(\tau)
   &=
   | \, C_{k}^{(+,-)}(\tau) \, |^2
   =
   | \, F_{0;k}^{(+)\,\dagger}(\tau) \,
        F_{k}^{(-)}(\tau) \, |^2
   \\
   &=
   1
   -
   | \, F_{0;k}^{(+)\,\dagger}(\tau) \,
        F_{k}^{(+)}(\tau) \, |^2
   \\
   &=
   1
   -
   \Tr{ \rho_{0;k}^{(+)} \, \rho_{k}^{(+)}(\tau) \, |^2 }
   \\
   &=
   \frac{1}{2} \,
   \bigl [ \,
      1 - \bP_{0;k}^{(+)} \cdot \bP_{k}^{(+)}(\tau) \,
   \bigr ] \>.
\end{split}
\end{equation}
We see immediately that $n_k(\tau_0) = 0$ at $\tau=\tau_0$.

We note that in the one-field method the current is automatically zero at $\tau=\tau_0$: Eq.~\eqref {bi.me.e:maxwellfinal} with
\begin{equation}
   P_3^{(+)}(\tau_0) = \frac{K_3(\tau_0)}{\omega_0}
                     = \frac{\pet}{\omega_0}
   \>,
\end{equation}
leads to a zero current because the integrand in Eq.~\eqref{bi.me.e:maxwellfinal} is odd in $\pet$. Furthermore, one of the subtleties of the one-field method is that the zero-current point is an unstable equilibrium point.  This is most easily seen from the equation of motion, Eq.~\eqref{bi.e:Peom}, of the polarization vector.  For $\tau = \tau_0$, we find that
\begin{equation}\label{bi.ic.e:dotPzero}
\begin{split}
   \partial_\tau \, \bP_k^{(+)}(\tau_0)
   &=
   2 \, \bK_k(\tau_0) \times \bP_k^{(+)}(\tau_0)
   \\
   &=
   2 \, \bK_k(\tau_0) \times \bK_k(\tau_0) / \omega_0
   =
   0 \>.
\end{split}
\end{equation}
However the second derivative is not zero:
\begin{equation}\label{bi.ic.e:ddotPzero}
   \partial_\tau^{2} \, \bP_k^{(+)}(\tau_0)
   =
   -
   \frac{ 2 m^2 }{ \sqrt{k^2 + 1} } \,
   \bigl ( \,
      k - e E_0 / m^2 \,
   \bigr ) \, \be_y \>.
\end{equation}

%
%%%%%%%%%%%%%%%%%%%%%%%%%%%%%%%%%%%%%%%%%%%%%%%%%%%%%%%%%%%%%%%%%%%%%%
%

\begin{figure}[t!]
   \centering
   \includegraphics[width=0.9\columnwidth]{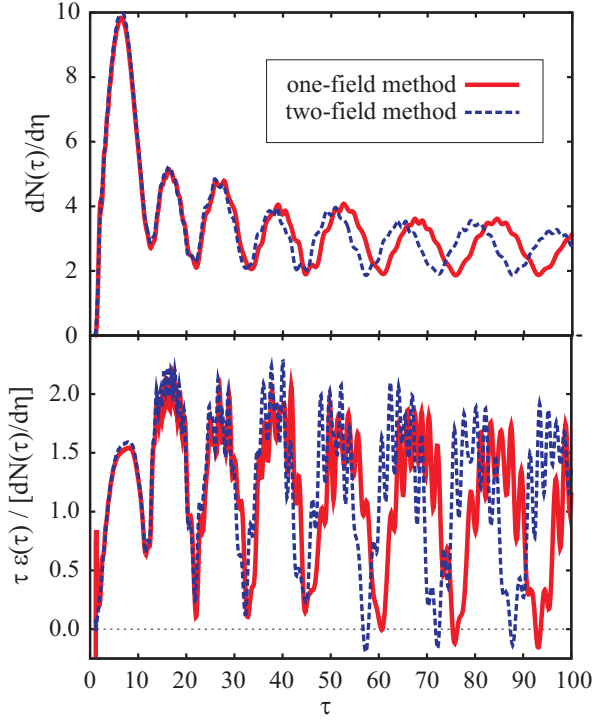}
   \caption{\label{fig3}(Color online)
   Proper-time evolution of the particle density,
   $\mathrm{d}N/\mathrm{d}\eta$ and proper time evolution of the ratio
   $\tau \varepsilon(\tau) / [ \mathrm{d}N/\mathrm{d}\eta]$ for the
   one-field and two-field methods.}
\end{figure}

\begin{figure*}[t!]
   \centering
   \includegraphics[width=0.7\textwidth]{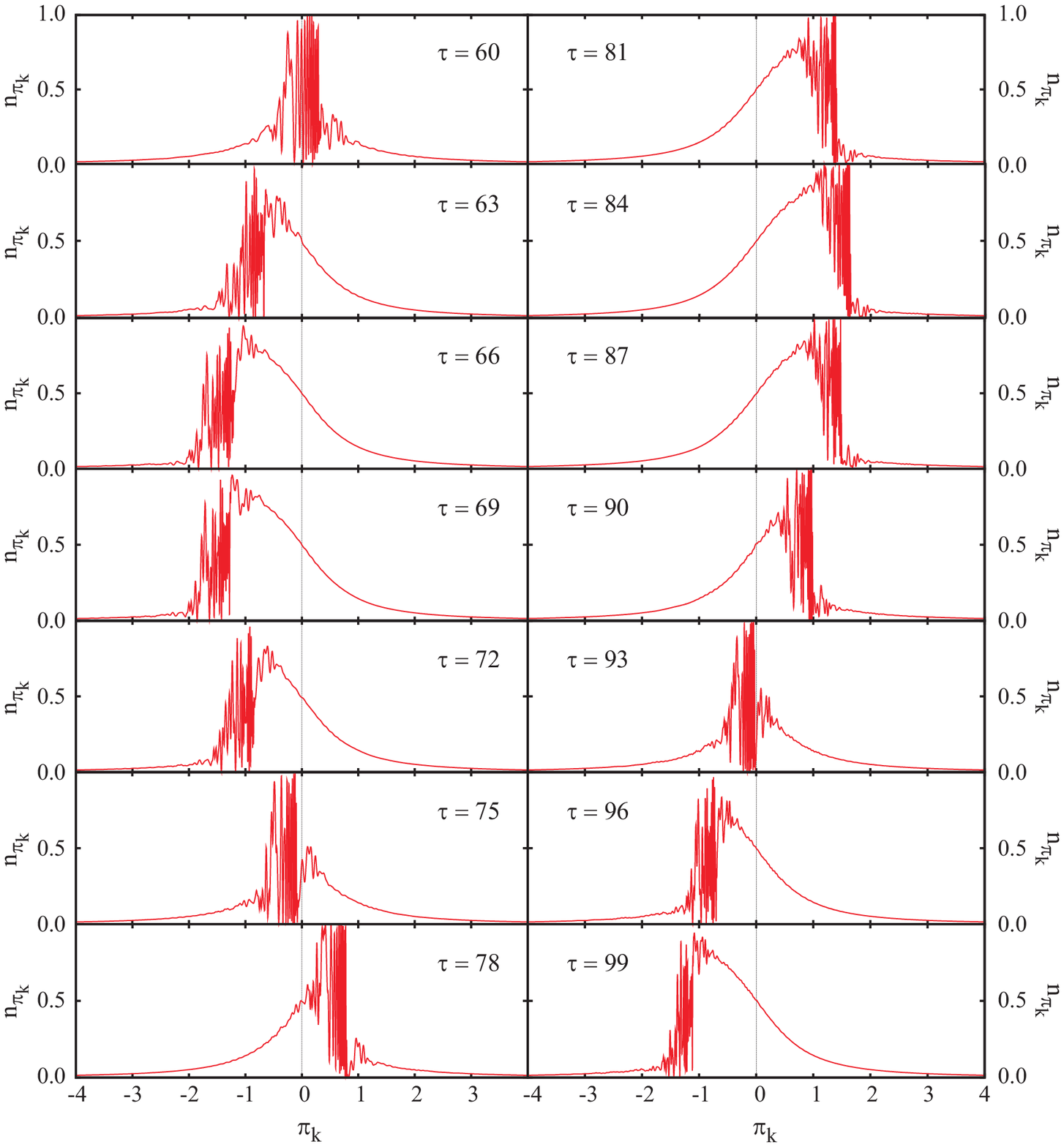}
   \caption{\label{fig4}(Color online)
   Proper-time evolution of the momentum dependent particle density
   distribution, $n_{\pet}$ defined in Eq.~\eqref{ic.e:ndef}, showing
   the oscillation of the centroid of the particle-density distribution
   between positive and negative values of $\pet$. Here, we show results
   for the one-field method.  Results for the two-field initial conditions
   scenario (not shown) are very similar, as it is to be expected from the
   results illustrated in Fig.~\ref{fig3}.}
\end{figure*}

%
%%%%%%%%%%%%%%%%%%%%%%%%%%%%%%%%%%%%%%%%%%%%%%%%%%%%%%%%%%%%%%%%%%%%%%
%

\subsection{Two field method}
\label{ss:twofield}

Here we start from solutions of the second-order Dirac equation.  Writing
the spinor $F_k(\tau)$ in the form:
\begin{equation}\label{bi.tf.e:Ffg}
   F_k^{(+)}(\tau)
   =
   \begin{pmatrix}
      f_{k,+}^{(+)}(\tau) \\
      f_{k,-}^{(+)}(\tau)
   \end{pmatrix} \>,
\end{equation}
from Dirac's Eq.~\eqref{bi.e:Fequ}, we can find a second-order equation for either the upper or lower component:
\begin{equation}\label{bi.tf.e:fdeII}
   \bigl \{ \,
      \partial_\tau^2
      +
      \omepi^2(\tau)
      -
      i \, s \,
      \dpet(\tau) \,
   \bigr \} \, f_{k,s}^{(+)}(\tau)
   =
   0 \>,
\end{equation}
where $s=\pm 1$ designates the upper or lower component.  A parametrization of these mode functions of the form:
\begin{multline}\label{bi.tf.e:fparaI}
   f_{k,s}^{(+)}(\tau)
   =
   \frac{ \mathcal{A}_{k,s}^{(+)} }
        { \sqrt{2 \, \Omega_{k,s}^{(+)}(\tau)} } \,
        \\ \times
   \exp
   \Biggl \{
      - i
      \int_{\tau_0}^{\tau}
      \Bigl [ \,
         \Omega_{k,s}^{(+)}(\tau')
         -
         s \,
         \frac{ i \dpet(\tau') }
              { 2 \, \Omega_{k,s}^{(+)}(\tau') } \,
       \Bigr ] \, \rd \tau'
    \Biggr \} \>,
\end{multline}
leads to a second-order nonlinear equation for $\Omega_{k,s}^{(+)}(\tau)$ given by:
\begin{equation}\label{bi.tf.e:nonlinear}
   \frac{1}{2}
   \frac{ \ddot{\Omega}_{s} }
        { \Omega_{s} }
   -
   \frac{3}{4}
   \biggl [
      \frac{ \dot{\Omega}_{s} }
           { \Omega_{s} }
   \biggr ]^2 \!\!\!
   +
   \frac{1}{2}
   \frac{ s \, \ddpet }
        { \Omega_{s} }
   -
   \frac{1}{4}
   \biggl [
      \frac{ \dpet }
           { \Omega_{s} }
   \biggr ]^2 \!\!\!
   -
   \frac{ s \,\dpet \,
           \dot{\Omega}_{s} }
        { \Omega_{s}^{2} }
   +
   \Omega_{s}^{2}
   =
   \omepi^2 \>.
\end{equation}
Here, and in the following, we suppress the dependencies on $\tau$, $k$, and the positive energy superscript.  Solutions of the nonlinear equation \eqref{bi.tf.e:nonlinear} for $\Omega_{s}$, subject to initial conditions given below, completely determine $f_{s}$.  Once we find $f_{s}$, we can get the other Dirac component from Dirac's equation:
\begin{equation}\label{bi.tf.e:fms}
   f_{-s}
   =
   \frac{1}{m} \,
   \bigl ( \,
      i \, \partial_\tau
      +
      s \, \pet \,
   \bigr ) \, f_{s}
   =
   \frac{ Z_{s}}{ m } \,
   f_{s} \>,
\end{equation}
where $Z_{k,s}^{(+)}(\tau)$ is given by:
\begin{equation}\label{bi.tf.e:Zdef}
   Z_{s}
   =
   X_{s}
   +
   i \, Y_{s}
   =
      \Omega_{s}
      +
      s \, \pet \,
   -
   i \
      \frac{
         \dot{\Omega}_{s}
         +
         s \, \dpet \,
           }{ 2 \, \Omega_{s} } \,
   \>,
\end{equation}
The normalization requirement: $ \sum_s|f_{s}|^2 = 1$ means that:
\begin{equation}\label{bi.tf.e:magf2pm}
   | \, f_{s} \, |^2
   =
   \frac{m^2}{ m^2 + | \, Z_{s} \, |^2 } \>,
   \quad
   | \, f_{-s} \, |^2
   =
   \frac{| \, Z_{s} \, |^2}
        { m^2 + | \, Z_{s} \, |^2 } \>,
\end{equation}
which fixes the normalization factor $\mathcal{A}_{s}$.  It is an easy matter now to get \emph{all} the terms of the density matrix $\rho_{s}$, and we find:
\begin{subequations}\label{bi.tf.e:pOmega}
\begin{align}
   P_{1;s}
   &=
   \frac{ 2 \, X_{s} }
        { m^2 + | \, Z_{s} \, |^2 } \>,
   \label{bi.tf.e:pxOmega} \\
   P_{2;s}
   &=
   \frac{ 2 \, Y_{s} }
        { m^2 + | \, Z_{s} \, |^2 } \>,
   \label{bi.tf.e:pyOmega} \\
   P_{3;s}
   &=
   \frac{ m^2 - | \, Z_{s} \, |^2 }
        { m^2 + | \, Z_{s} \, |^2 } \>.
   \label{bi.tf.e:pzOmega}
\end{align}
\end{subequations}

We are now in a position to carry out an adiabatic expansion of the nonlinear equation \eqref{bi.tf.e:nonlinear}.  We again count derivatives with respect to $\tau$ by putting:
$\partial_\tau \mapsto \epsilon \, \partial_\tau$, and expand
\begin{equation}\label{bi.tf.e:Omegaexp}
   \Omega_{s}
   =
   \Omega^{(0)}_{s}
   +
   \epsilon \, \Omega^{(1)}_{s}
   +
   \epsilon^2 \, \Omega^{(2)}_{s}
   +
   \dotsb
\end{equation}
Inserting this into \eqref{bi.tf.e:nonlinear} and inverting the equation gives $\Omega^{(0)}_{s} = \omepi$ and $\Omega^{(1)}_{s} = 0$, from which we find:
\begin{equation}\label{bi.tf.e:Omega2}
   \Omega^{(2)}_{s}
   =
   \frac{\omepi - s \, \pet }{ 2 \, \omepi } \,
   \Bigl [ \,
   \frac{1}{2} \,
   \frac{ s \, \ddpet }{ \omepi^2 } \,
   +
   \frac{ \dpet^2 }{ \omepi^3 } \,
   -
   \frac{5}{4} \,
   \frac{ \dpet^2 }{ \omepi^4 } \,
   ( \, \omepi + s \, \pet \, ) \,
   \Bigr ] \>.
\end{equation}
From this we find that
\begin{align}
   Z_{s}
   &=
   X_s + i \, Y_s
   \label{bi.tf.e:Zexp} \\
   &=
   ( \, \omepi - s \, \pet \, ) \,
   \Bigl [ \,
      1
      +
      i \epsilon \,
      \frac{s \, \dpet}{2 \, \omepi^2 }
      +
      \epsilon^2 \,
      \frac{ \Omega_{s}^{(2)} }
        { ( \, \omepi - s \, \pet \, ) }
      +
      \dotsb
   \Bigr ]
   \>.
   \notag
\end{align}
So from our general expressions \eqref{bi.tf.e:pOmega}, it is easy to show that:
\begin{align}
   P_{1;s}
   &=
   \frac{m}{\omepi}
   +
   \epsilon^2 \, m \,
   \Bigl ( \,
      -
      \frac{1}{8} \,
      \frac{\dpet^2}{\omepi^5}+
      \frac{1}{4} \,
      \frac{ \pet \, \ddpet}{\omepi^5}
      -
      \frac{5}{8} \,
      \frac{\pet^2 \, \dpet^2}{\omepi^7} \,
   \Bigr )
   +
   \dotsb
   \notag \\
   P_{2;s}
   &=
   \epsilon \, m \,
   \frac{s \, \dpet}{2 \, \omepi^3}
   +
   \dotsb
   \label{eq:Pzs} \\
   P_{3;s}
   &=
   \frac{s \, \pet}{\omepi}
   -
   \epsilon^2 \, s \, m^2 \,
   \Bigl ( \,
      \frac{1}{4} \,
      \frac{ \ddpet }{ \omepi^5 }
      -
      \frac{5}{8} \,
      \frac{ \pet \, \dpet^2 }{ \omepi^7 } \,
   \Bigr )
   +
   \dotsb
   \notag
\end{align}
For $s=1$, Eqs.~\eqref{eq:Pzs} are in agreement with Eqs.~\eqref{bi.ae.e:PxPyPz}.
So to second adiabatic order $P_{1;s}$ is independent of $s$, but $P_{2;s}$ and $P_{3;s}$ change sign with $s$.

To specify the initial conditions for second-order nonlinear Eq.~\eqref{bi.tf.e:nonlinear} at $\tau = \tau_0 = 1/m$ one needs two initial conditions. Since the vacuum state is not unique when particles are being produced, one usually chooses some approximate adiabatic vacuum state of given order as discussed in Ref.~\onlinecite{r:BirrellDavies}.  The authors in Ref.~\onlinecite{r:Cooper:1993uq} chose the first-order adiabatic conditions as
\begin{subequations}\label{bi.tf.e:yuvalinitial}
\begin{align}
   \Omega_{k,s}^{(+)}(\tau_0)
   &=
   \omega_0
   =
   m \, \sqrt{k^2 + 1} \>,
   \\
   \dot{\Omega}_{k,s}^{(+)}(\tau_0)
   &=
   \dot \omega_0
   =
      m^2 \,
   \frac{ k \, ( \, \tilde{E}_0 - k \, ) }
        { \sqrt{ k^2 + 1 } } \>,
\end{align}
\end{subequations}
where $\tilde{E}_0 = e E_0 / m^2$.  The initial conditions are independent of $s$.

Using Eq.~\eqref{eq:Pzs}, we obtain that for each value of $s$ this choice of initial conditions at $\tau = \tau_0$ will lead to a non-vanishing current,~$J_{s}(\tau_0)$. However, if we average over the two sets of solutions $s= \pm$ and chooses for the Maxwell equation:
\begin{equation}\label{bi.me.e:maxwellfinal2}
   \partial_\tau E(\tau)
   =
   \frac{e}{2}
   \Intk{\pet} \left[ P_{3,+}^{(+)}(\pet,\tau) +  P_{3,-}^{(+)}(\pet,\tau )\> \right].
\end{equation}
then the renormalized Maxwell equation will start with a zero value for the current.

%
%%%%%%%%%%%%%%%%%%%%%%%%%%%%%%%%%%%%%%%%%%%%%%%%%%%%%%%%%%%%%%%%%%%%%%
%

\section{Numerical results}
\label{s:results}

We have performed numerical calculations for both sets of initial conditions described above. We employed a fourth-order Runge-Kutta method to solve the coupled Dirac equation and backreaction problem.  The $k$-momentum variable, which is dimensionless, was discretized on a nonuniform piece-wise momentum grid with a cutoff at $k = \Lambda_k$.  We found that a value of $\Lambda_k \approx 200$ was necessary to obtain numerical results insensitive with respect to the cutoff.  For the purpose of calculating the subtracted values of the current $J(\tau)$, matter energy $\varepsilon(\tau)$, matter pressure $p(\tau)$, and fermion particle density $\mathrm{d}N(\tau)/\mathrm{d}\eta$, we needed to compute the momentum integrals with respect to the variable, $\pet$ rather than $k$.  The corresponding momentum cutoff in $\pet$-space was chosen to be 20\% greater than $\tau_\mathrm{max} \Lambda_k$ to allow for possible very large values of $A(\tau)$, which is unknown at the beginning of the calculation.  The momentum integrals in $\pet$-space were performed using a Chebyshev integration method with spectral convergence~\cite{r:MM02}. Using the procedure outlined here, we found that approximately 8000 mode functions were necessary to obtain a converged numerical result.  The conservation of the energy-momentum tensor, see Eq.~\eqref{bi.em.e:consI}, served as a numerical test: we found that the renormalized energy-momentum tensor was conserved within machine precision.

For the purpose of this comparison, we took: $m=1$, $e=1$, $\tau_0 = 1/m = 1$, $A(\tau_0)=0$, and $E(\tau_0)=4$.  These strong-field initial conditions have been shown to produce sufficient fermion pairs at $\tau = \tau_0$ for plasma oscillations to take place.  In Fig.~\ref{fig1}, we show the proper-time evolution of the electromagnetic field, $A(\tau)$, electric field, $E(\tau)$, and current, $J(\tau)$, for the one-field and two-field methods described in text.  The components of the matter part of the energy-momentum tensor, $\varepsilon(\tau)$ and $p(\tau)$, for the two simulations are shown in Fig.~\ref{fig2}.  Finally, the proper-time evolution of the particle density, $\mathrm{d}N/\mathrm{d}\eta$, defined in Eq.~\eqref{ic.e:ndef}, is given in Fig.~\ref{fig3}.  For both methods, the ratio $\tau \varepsilon(\tau) / [ \mathrm{d}N/\mathrm{d}\eta]$ is seen to oscillate around the numerical value of 1, consistent with the hydrodynamical picture, as explained in Ref.~\onlinecite{r:Cooper:1993uq}.  We notice that the two sets of solutions are almost identical at short and intermediate times.  The two solutions become out of phase at late times due to the slightly different initial conditions.  However, in the real problem we expect that interactions between the fermions would eliminate these oscillations.

The proper-time evolution of the momentum-dependent particle-density distribution, $n_{\pet}$, corresponding to the choice of initial conditions in the one-field method, is shown in Fig.~\ref{fig4}. We note that the centroid of the particle-density distribution oscillates between positive and negative values of $\pet$. The oscillation of the number density is a result of the current oscillating in sign, the current in momentum space being related to the number density times the velocity of light.  This effect is also seen classically when two infinite oppositely charged parallel plates initially a finite distance apart are released and allowed to pass through one another. In that case both the current and electric field oscillate in an analytically derivable manner~\cite{r:CB1989sf}. Results for the case of the two-field method (not shown) are very similar, as it is to be expected from the results depicted in Fig.~\ref{fig3} (see also Ref.~\onlinecite{ref:qed1dmovies}).

%
%%%%%%%%%%%%%%%%%%%%%%%%%%%%%%%%%%%%%%%%%%%%%%%%%%%%%%%%%%%%%%%%%%%%%%
%

\section{Conclusions}
\label{s:conclusions}

To conclude, in this paper we report an initial-conditions sensitivity study for the problem of pair production of fermions coupled to a ``classical''  electromagnetic field with backreaction in \oneplusone\ boost-invariant coordinates. We discuss two methods of choosing the initial conditions which are consistent with having the fermions in a  ``vacuum state.'' We conclude that the two methods of starting out the calculation produce essentially the same answer. Based on our numerical simulations, there seems to be little reason theoretically or otherwise to use the two-field method discussed previously in Ref.~\onlinecite{r:Cooper:1993uq}, as it doubles the storage requirements and computational time. This is important for our forthcoming studies of fermion particle production with backreaction in QED and QCD.

We emphasize here that in the case of the squared Dirac equation (two-field method) there are two independent solutions of the second-order differential equation for the mode functions, each of which provides a basis for two different fermi fields.  In order to make compatible the physical requirement that the initial current is zero with the initial choice that the fermions were initially chosen to be a first-order adiabatic vacuum state, the authors of Ref.~\onlinecite{r:Cooper:1993uq} simply \emph{averaged} these two solutions to produce a current which was zero at $\tau = \tau_0$.  So doubling the number of  fermi fields allows one to produces consistent initial conditions if we define the current by averaging over the two sets of solutions.  By staying with the original first-order Dirac equation, in the one-field method we were able to satisfy the initial condition of zero current by choosing a slightly cruder initial state for the fermion fields. This choice, however, reduces by half the size and duration of the calculation.

%
%%%%%%%%%%%%%%%%%%%%%%%%%%%%%%%%%%%%%%%%%%%%%%%%%%%%%%%%%%%%%%%%%%%%%%
%

\begin{acknowledgments}
This work was performed in part under the auspices of the United States Department of Energy. The authors would like to thank the Santa Fe Institute for its hospitality during the completion of this work.
\end{acknowledgments}

%
%%%%%%%%%%%%%%%%%%%%%%%%%%%%%%%%%%%%%%%%%%%%%%%%%%%%%%%%%%%%%%%%%%%%%%
%

\bibliography{johns}

%
%%%%%%%%%%%%%%%%%%%%%%%%%%%%%%%%%%%%%%%%%%%%%%%%%%%%%%%%%%%%%%%%%%%%%%
%

\end{document}